\documentclass[
journal=nalefd, 
manuscript=letter,
layout=twocolumn,
doi=true,
]{achemso}
\usepackage[utf8]{inputenc} % set input encoding (not needed with XeLaTeX)

\usepackage{graphicx} % support the \includegraphics command and options

\usepackage{cmap}
\usepackage[T1]{fontenc}
\usepackage{booktabs} % for much better looking tables
\usepackage{array} % for better arrays (eg matrices) in maths
\usepackage{paralist} % very flexible & customisable lists (eg. enumerate/itemize, etc.)
\usepackage{verbatim} % adds environment for commenting out blocks of text & for better verbatim
\usepackage{amsmath}
\usepackage{textcomp}
\usepackage[english]{babel}
\usepackage{xcolor}

\newcommand{\latin}[1]{\textit{#1}}

\newcommand*{\unit}[2]{{#1}\,{\ensuremath{\mathrm{#2}}}}
\newcommand*{\ang}[1]{#1\text{\textdegree}}
\newcommand*{\um}{\text{\textmu}{m}}
\newcommand*{\angstrom}{\ensuremath{\text{\AA}}}
\makeatletter
\newcommand*{\@IonType}{}
\newcommand*{\@IonHelper}[1][+]{\@IonType\textsuperscript{#1}\endgroup}
\newcommand*{\ion}[1]{\begingroup\renewcommand*{\@IonType}{#1}\@IonHelper\@IonType}
\makeatother
\newcommand{\E}[1]{\ensuremath{\times 10^{#1}}}

\newcommand{\degreeC}{\text{\textcelsius}}

\title{Nano patterns self-aligned to Ga dimer rows on GaAs surfaces}

\author{Martin {Engler}}
\email{m.engler@hzdr.de}
\affiliation{
	Helmholtz-Zentrum Dresden-Rossendorf, 
	Institute of Ion Beam Physics and Materials Research, 
	Bautzner Landstra\ss{}e 400, 
	01328 Dresden, 
	Germany}
\author{Tom\'a\v{s} {\v{S}kere\v{n}}}
\affiliation{
	IBM Research -- Zurich, 
	S\"aumerstrasse 4, 
	8803 R\"uschlikon, 
	Switzerland}
\alsoaffiliation{Czech Technical University in Prague, 
	Faculty of Nuclear Sciences and Physical Engineering, 
	B\v{r}ehov\'a 7, 
	115 19 Prague 1, 
	Czech Republic}
\author{Stefan {Facsko}}
\affiliation{
	Helmholtz-Zentrum Dresden-Rossendorf, 
	Institute of Ion Beam Physics and Materials Research, 
	Bautzner Landstra\ss{}e 400, 
	01328 Dresden, 
	Germany}

\begin{document}
\begin{abstract}
	Ion beam irradiation of semiconductors is a method to produce regular periodic nanoscale patterns self-organized on wafer scale. 
	At low temperatures, the surface of semiconductors is typically amorphized by the ion beam. 
	Above a material dependent dynamic recrystallization temperature however, the surface remains crystalline and ion beam irradiation produces regular arrays of faceted ripples on III-V semiconductors.
	This provides a powerful single-step technique for the production of nanostructures on a large area. 
	On $(001)$ surfaces these ripples are parallel to the $[1\bar{1}0]$ direction without any external anisotropy. 
	The origin of this self-alignment was not fully understood until now. 
	A simple experiment exposing the front side and back side of a GaAs$(001)$ wafer to the ion beam clarifies its origin and proves that the ripples align to the Ga dimer rows.
	As the direction of Ga dimer rows rotates by 90\textdegree{} on the back side, the orientation of the ripples also rotates by 90\textdegree{} to $[110]$. 
	We discuss the experimental results in view of a model where the pattern formation is driven by creation of vacancies and ad-atoms by the ion beam and their diffusion, which is linked to the direction of dimers on the surface.%
\end{abstract}
\begin{tocentry}
	\centering
	\includegraphics{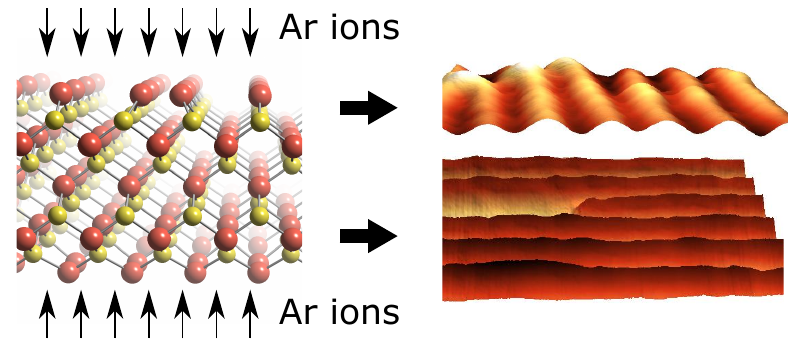}
\end{tocentry}
	
%\pacs{
%	81.16.Rf, %Micro- and nanoscale pattern formation
%	68.37.Ps, %Atomic force microscopy (AFM)
%	61.80.Jh, %Ion radiation effects
%	81.05.Ea, %III-V semiconductors
%%	81.65.Cf %Surface cleaning, etching, patterning
%}

\maketitle

\vspace{2\baselineskip}
Nanoscale patterning of surfaces is of crucial importance for future applications e.\,g. in optoelectronics \cite{Wang2013, Zhang2014, Atwater2010, Catrysse2010, Wierer2009}, plasmonics \cite{Atwater2010, Lee2013}, and medical diagnosis \cite{Lee2013}. In these applications regular, well ordered patterns on a large scale are essential. Simple, scalable methods allowing the treatment of large areas are required for their production. Self-organized ion beam patterning is such a method, which can pattern surfaces on a wafer-scale. It is simple and versatile and can be used to produce ripple patterns \cite{Munoz-Garcia2014, Chowdhury2015, Ou2015, Teichmann2013, Engler2014a}, hexagonal dot patterns \cite{El-Atwani2015, Facsko1999,  Munoz-Garcia2014, Munoz-Garcia2010, Frost2000}, or faceted periodic structures \cite{Engler2014a, Teichmann2013, Basu2013, Ou2015, Chowdhury2015} with periodicities from \unit{10}{nm} to several \unit{100}{nm}. At room temperature the ion beam amorphizes the surface \cite{Engler2016}.  Ion beam patterning at elevated temperatures above a material dependent dynamic recrystallization temperature $T_\text{c}$, however, yields crystalline periodic patterns \cite{Chowdhury2015, Ou2013, Ou2015, Engler2016}.
These crystalline patterns can be used to grow periodic metal nanostructures epitaxially without using lithography \cite{Monteverde2001}.

%\Todo{Verweis auf Musterbilung auf Metallen}
While below $T_\text{c}$ ion beam pattern formation of III-V semiconductors is driven by composition modulations \cite{El-Atwani2015, Bradley2010, Bradley2012, Norris2014, Shenoy2007}, ion beam pattern formation above $T_\text{c}$ is dominated by surface defect creation and diffusion \cite{Ou2013, Ou2015, Hashmi2016, Renedo2016}.
Above $T_\text{c}$ extremely ordered faceted ripples form on GaAs(001) as recently reported \cite{Chowdhury2015, Ou2015}.
Their orientation is assumed to be linked to the direction of dimer rows on the surface \cite{Ou2015, Chowdhury2015, Hashmi2016},
however, conclusive experimental evidence is still missing.
From molecular beam epitaxy (MBE) it is known that the Ehrlich-Schwoebel (ES) barrier at step-edges for the deposited adatoms destabilizes the surface \cite{Villain1991}. The diffusing species are the deposited adatoms in MBE. On ion beam irradiated surfaces the situation is similar: Ion impacts produce mobile surface vacancies and adatoms.
The vacancies and adatoms diffusing on the surface are hindered by an ES barrier to cross step edges. This induces an effective non-equilibrium vacancy current from hill tops to valleys and an adatom current from the valleys to the hill tops. The vacancy current is effectively the same as a mass current in the opposite direction. So both currents add to an effective uphill mass current. 
These similarities with MBE lead Ou~\latin{et~al.} \cite{Ou2013} to naming this pattern formation process ``reverse epitaxy''.
With increasing slope, the collision cascade will transport more atoms downhill by ballistic mass drift, as proposed by Carter and Vishnyakov \cite{Carter1996}. Also the random nature of nucleation of vacancy and adatom islands on the terraces will smooth surface fluctuations, especially on short length scales \cite{Politi2000}. The interplay of these mechanisms determines the  pattern characteristics, like wavelength and facet slopes and their temporal evolution.

\begin{figure}
	\centering
	\includegraphics{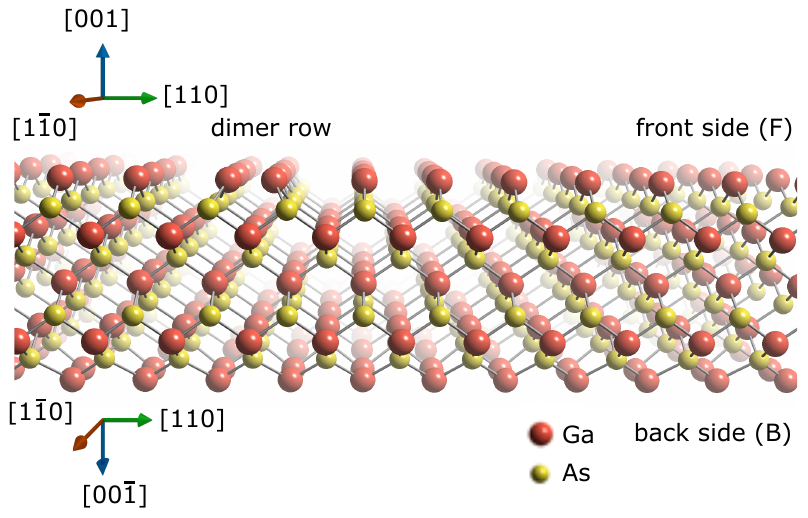}
	\caption{\label{fig:structure}
		Structure models of  GaAs(001) (F) and  GaAs$(00\bar{1})$ (B) surfaces.
		}
\end{figure}
We present a model explaining the alignment of the ripples to this direction.
Due to the Zincblende crystal structure, GaAs(001) is a highly anisotropic surface (Fig.~\ref{fig:structure}) with only two-fold symmetry. 
The possible surface reconstructions at the experimental conditions form dimer rows in $[1\bar{1}0]$ direction \cite{Ohtake2008}.
This low symmetry leads to an anisotropic diffusion which is much faster along $[1\bar{1}0]$ than along $[110]$ \cite{Salmi1999}. 
Also the energy of steps is highly anisotropic: The energy cost for creating steps parallel to $[1\bar{1}0]$ is much lower than for steps parallel to $[110]$ \cite{Magri2014}. These anisotropies lead to the nucleation of vacancy and adatom islands elongated in $[1\bar{1}0]$ starting the formation of ripples which then align to $[1 \bar{1} 0]$.
Our model also explains the facet angles deviating from thermodynamically stable facets.

This model can be tested by a simple experiment, where two samples cut from the same wafer exposing the front side (F), i.\,e.GaAs$(001)$, and the back side (B), i.\,e. GaAs$(00\bar{1})$, of the wafer, respectively, to the ion beam are irradiated simultaneously.
Due to the symmetry of the Zincblende structure, dimer rows, the low energy steps and the fast diffusion direction are parallel to $[110]$ on the (B) side, i.\,e. GaAs$(00\bar{1})$.
Thus, the direction of the ripples is expected to rotate by \ang{90} on GaAs$(00\bar{1})$ (B) compared to GaAs(001) (F) (see Fig.~\ref{fig:structure}). 

For the experiment, we cut two $\unit{5}{mm}\times\unit{10}{mm}$ samples from the \emph{same} double side polished GaAs(100) wafer.  Both samples were mounted on the same sample holder. One sample (F) exposed the front side and the other sample (B) the back side  of the wafer, corresponding to the surface orientations $(001)$ and $(00\bar{1})$ respectively, to the ion beam. Both samples were irradiated at the same time with  \unit{1}{keV} \ion{Ar} at normal incidence and at a temperature of \unit{370}{\degreeC}. We used a broad beam Kaufman type ion source, which provides a homogeneous flux of $J=\unit{1\E{15}}{cm^{-2}\,s^{-1}}$ on both samples. The fluence was \unit{5\E{18}}{cm^{-2}}. The temperature was measured using an infrared pyrometer. The resulting patterns were analyzed ex-situ with atomic force microscopy (AFM) in tapping mode. The depth of the sputter crater was measured with a stylus profilometer. For imaging with scanning tunneling microscopy (STM) a GaAs(001) sample was irradiated with a fluence of \unit{5\E{18}}{cm^{-2}} at \unit{450}{\degreeC}, transferred to an ultra high vacuum STM system, and cleaned in-situ with \unit{2\E{15}}{cm^{-2}} \unit{1}{keV} \ion{Ar} at \unit{450}{\degreeC}. The AFM and STM images were analyzed using the software \textsc{Gwyddion}\cite{Necas2012}.

%\section{Results + Discussion}
\begin{figure*}
	\centering
	\includegraphics{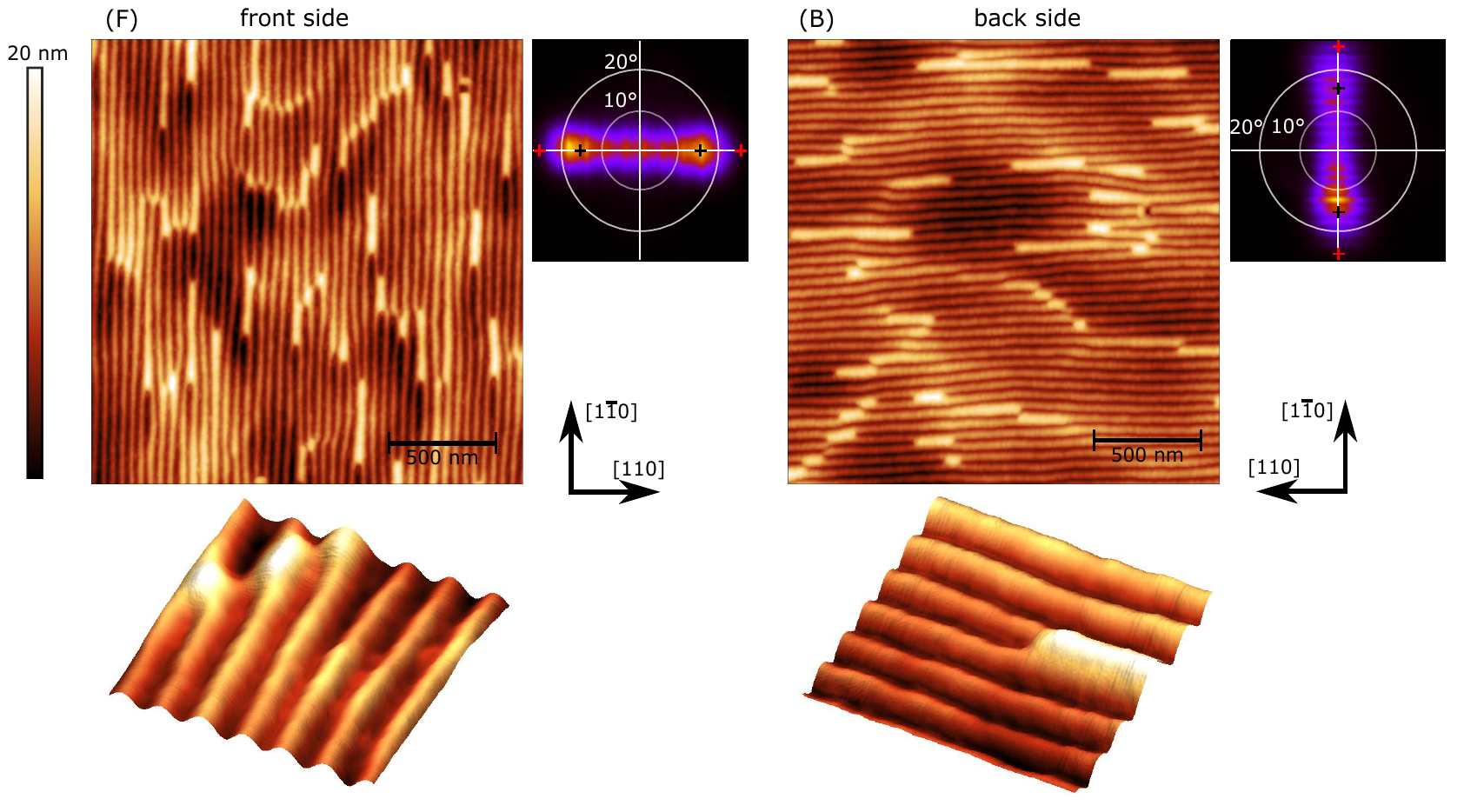}
	\caption{\label{fig:front_back_AFM}
		AFM topographs of irradiated front side (F) and back side (B) of the wafer.
		For each sample also a detailed 3D image with \unit{$300\times300$}{nm^2} image size and the slope distribution are shown, where the red and black crosses mark the $\{113\}$ and $\{115\}$ facets, respectively.
		}
\end{figure*}

First we will discuss the simultaneous irradiation of the front side GaAs(001) (F) and back side GaAs$(00\bar{1})$ (B) of one wafer. Fig.~\ref{fig:front_back_AFM} shows an AFM topography, the corresponding slope distribution, and a 3D detail image for each sample.
The AFM topographs clearly show that on sample (F) the ripples are parallel to the $[1\bar{1}0]$ direction, while on sample (B) they are parallel to the $[110]$ direction. Both samples exhibit almost the same ripple wavelength of \unit{48}{nm} and \unit{49}{nm} for sample (F) and (B), respectively. Also the surface roughness is almost the same with \unit{3.3}{nm} and \unit{3.5}{nm}, respectively. The slope distributions are dumbbell shaped and aligned with $[110]$ and $[1\bar{1}0]$, respectively. They have broad peaks at $\approx \ang{16}$, close to the $\{115\}$ facets with a slope of \ang{15.8}, marked with black crosses. The largest slopes present on the surface are at \ang{25}, close to the $\{113\}$ facets with a slope of \ang{25.5}, marked with red crosses.
The ripples created at \unit{370}{\degreeC} are less ordered than the ripples produced by Ou \latin{et~al.} \cite{Ou2015} at \unit{410}{\degreeC} and exhibit bifurcations. The bifurcations are shown in detail in the 3D magnifications, which show that  the slope of the facets remains constant at the bifurcations.
The direction of the ripples rotates together with the direction of the Ga atomic rows on the surface. At normal incidence the ion beam does not induce any anisotropy, so the only anisotropy in the system is the crystal structure of the surface. We have seen that the orientation of the ripple pattern is directly linked to the direction of the atomic rows on the surface and not to a specific crystallographic direction.

%\subsection{Model}
In the following we will discuss the mechanisms of pattern formation on the GaAs(001) surface (F). On the GaAs$(00\bar{1})$ surface (B), the directions $[1\bar{1}0]$ and $[110]$ have to be exchanged.
During an ion impact, the collision cascade mixes atoms, sputters atoms and locally heats the surface close to the melting temperature for a few ps. Above $T_{\text{c}}$, the heat deposited by the impact provides enough mobility to heal the disorder. After the impact site has cooled, the crystal lattice is restored and only a few point defects remain at the surface as adatoms and surface vacancies, which diffuse on the surface. As the crystal structure is restored after an ion impact and no amorphization takes place, we assume that the reconstruction is  restored as well, leading to anisotropic diffusion.
As is preferentially sputtered \footnote{The partial sputtering yields, calculated with SRIM 2013 (\url{www.srim.org}) are $Y_{\text{As}}=2.7$ for As and $Y_{\text{Ga}}=1.2$ for Ga on a stoichiometric surface.}, so Ga will be enriched at the surface. 
Ga rich surfaces typically form reconstructions with dimer rows parallel $[1\bar{1}0]$ \cite{Such2003, Ohtake2008} below \unit{600}{\degreeC}, so it is reasonable to assume that the fast diffusion is here in the $[1\bar{1}0]$ direction.
On larger terraces the diffusing vacancies and adatoms nucleate to islands.  The energy cost for producing a step edge $\parallel [1\bar{1}0]$ is $\approx \unit{120}{meV/\angstrom}$ smaller than for a step edge $\parallel [110]$, according to LDA-DFT calculations \cite{Magri2014}. 
The attachment to existing kinks will not cost additional energy in this simple picture. The formation of a new kink however includes the formation of two short steps perpendicular to the original step. Thus, the energy cost for a new kink is much smaller at $[110]$ than at $[1\bar{1}0]$ steps. 
Adatoms and vacancies reaching a step diffuse along the step edge until they are incorporated into a kink. To reach a kink site they can diffuse around corners, so the islands grow faster in $[1\bar{1}0]$ than in $[110]$. These long islands start the ripple formation process. 
For simplicity, we will assume translational invariance in $[1\bar{1}0]$ and discuss the formation of perfect defect free ripples. In the following the x-axis is parallel to $[110]$.

Once the islands have formed, vacancies and adatoms will be preferentially incorporated into descending and ascending steps, respectively, due to the ES barrier. This leads to preferential nucleation of new vacancy islands on bottom and adatom islands on top terraces as the surrounding steps are not efficient sinks. This can be seen as an effective uphill mass current destabilizing the surface \cite{Politi2000}, which is given by 
\begin{equation}
	j_{\text{ES}} = F_0 \frac{
			l_{\text{ES}} l_{\text{D}}
			}{
			2(1+l_{\text{ES}}/l_{\text{D}})
			} m,
\end{equation} 
for small slopes $m=\partial_x h$.
The adatom/vacancy creation flux $F_0 = J(Y_{\text{ad}}+Y_{\text{vac}})$ depends on the ion flux $J$, the adatom $Y_{\text{ad}}$ and vacancy production yields $Y_{\text{vac}}$ of an ion impact.
The diffusion length on a terrace is given by $l_{\text{D}} \approx (D/F_0)^{1/2(d+1)}$ for $d$ dimensions, with the diffusivity $D$.
The Ehrlich-Schwoebel length $l_{\text{ES}} = D/(a \nu)$ increases with a stronger ES barrier. Here $a$ is the surface lattice constant and $\nu$ the rate to jump over the ES barrier.
The peak to valley height $<\unit{20}{nm}$ of the ripples is much smaller than the $\approx \unit{6.5}{\um}$ thickness of eroded material. A similar situation arises in homoepitaxy with a weak ES barrier, i.e. $l_{\text{ES}} \ll l_{\text{D}}$: The height of the mounds is much smaller than the thickness of the deposited layer \cite{Michely_Krug_2004}. In analogy we conclude that the ES barrier is weak here, too.

The ion beam also directly induces a mass current on the surface.
In the collision cascade atoms receive momentum in direction of the ion beam and adatoms are created further in direction of the ion beam than vacancies. This induces a downhill current for normal incidence proportional to the slope $m$ \cite{Carter1996,Norris2014}.
Also the emission of atoms from steps is larger than from terraces. Some of these atoms will slide on the lower terrace and have enough kinetic energy to cross the ES barrier. As the step density is proportional to the slope $m$, this current is proportional to $m$. So both mechanisms contribute to the ion induced current
\begin{equation}
j_{\text{ion}} \approx -J \alpha_{\text{ion}} m,
\end{equation}
where $\alpha_{\text{ion}}$ is a proportionality factor giving the strength of the ion induced current.

%\end{figure} 
The barrier of \unit{1.6}{eV} for detachment of adatoms bound to a step is so high that the rate of detachment is only $\approx\unit{3}{s^{-1}}$ at \unit{370}{\degreeC} \cite{Kazantsev2015}. 
The rate for detachment from kink sites is even two orders of magnitude lower. These rates are insignificant compared to adatom and vacancy creation by the ion beam with a flux $J=\unit{10}{nm^{-2}\,s^{-1}}$. Thus, thermal detachment of adatoms and vacancies can be neglected and classic Herring-Mullins diffusion is not operative here. 
The stochastic nature of nucleation on  terraces leads to an effective Herring-Mullins like current \cite{Politi2000}, which is
\begin{equation}
	j_{\text{nuc}} = F_0 l_{\text{D}}^4 \partial_x^3 h
\end{equation}
for small slopes. Assuming a weak ES barrier, i.e. $l_{\text{ES}} \ll l_{\text{D}}$, the evolution of the surface for small slopes is described by a continuum equation in linear approximation
\begin{equation}
\begin{split}
\partial_t h =& -V_{\text{a}} \partial_x (j_{\text{ES}} + j_{\text{ion}} + j_{\text{nuc}}) \\
=&\ V_{\text{a}} \bigg[
- \bigg(
\frac{F_0 l_{\text{ES}} l_{\text{D}} } {2} 
- J \alpha_{\text{ion}}
\bigg) \partial_x^2 h %\\
-  F_0 l_{\text{D}}^4 \partial_x^4 h
\bigg],
\end{split}
\end{equation}
where $V_{\text{a}}$ is the atomic volume.
The wavelength of the fastest growing Fourier mode 
\begin{equation}
	\lambda^* = 4\pi \sqrt{ 
		\frac{
			(Y_{\text{ad}}+Y_{\text{vac}}) l_{{D}}^4
		}{
			(Y_{\text{ad}}+Y_{\text{vac}}) l_{\text{D}} l_{\text{ES}} - 2 \alpha_{\text{ion}} 
		} 
	}
\end{equation}
depends only weakly on the flux, as $l_{\text{D}} \propto J^{-1/4}$  for a 1 dimensional system \cite{Politi2000}.
No flux dependence has been observed experimentally for reverse epitaxy of Si(001) \cite{Engler2016} and GaAs(001) \cite{Chowdhury2016} when the flux is changed by a factor of $\sim 4$.

\begin{figure}
	\centering
	\includegraphics{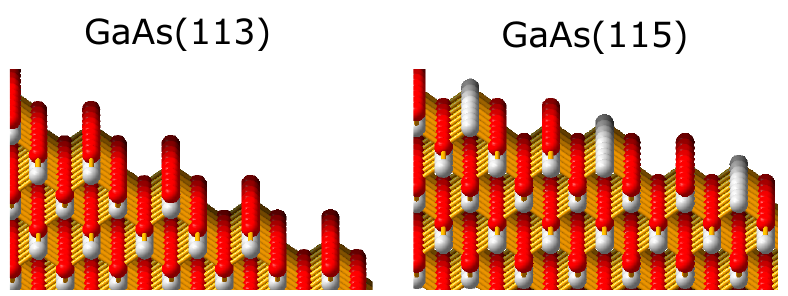}
	\caption{\label{fig:facet}
		Models of the unreconstructed \{113\} and \{115\} facets.
	}
\end{figure}
Next we will discuss the slope selection occurring for larger slopes.
When the local facet  approaches a high symmetry surface orientation the ES current is expected to vanish, as the surface cannot be seen anymore as a vicinal surface, i.e. the surface is no longer composed of (001) terraces and steps \cite{Siegert1994}.
 This can be modeled by 
\begin{equation}
	j_{\text{ES}} = J \alpha_{\text{ES}}\, m (1 - m^2/m_0^2),
\end{equation}
with 
$\alpha_{\text{ES}} = (Y_{\text{ad}}+Y_{\text{vac}})\frac{
	l_{\text{ES}} l_{\text{D}}
}{
2(1+l_{\text{ES}}/l_{\text{D}})
}$ 
with the facet slope  $m=\partial_x h$ and the slope $m_0$ of the thermodynamically stable low index surface \cite{Ou2013, Ou2015, Siegert1994}, which are the \{113\} facets with a slope angle of \ang{25.2} for GaAs(001) \cite{Usui2008}. 
 The sum of the ES current and the ion induced current 
\begin{equation}
	j_{\text{ES}} + j_{\text{ion}} 
	 = J (\alpha_{\text{ES}} + \alpha_{\text{ion}}) m - J \alpha_{\text{ES}} \frac{m^3}{m_0^2}
\end{equation}
vanishes on facets with a stable slope. So the selected slope is given by
\begin{equation}
	m^* = m_0 \sqrt{1 - \frac{\alpha_{\text{ion}}}{\alpha_{\text{ES}}}}\ .
\end{equation}
The ion beam reduces the slope of the stable facet compared to the low index facet. This relation can also be used to estimate the relative strength $\alpha_{\text{ion}} / \alpha_{\text{ES}}$ of the ion beam effects to the effect of the ES barrier. For $\alpha_{\text{ion}} / \alpha_{\text{ES}}=0$, there are no ion induced currents contributing to the surface evolution and for $\alpha_{\text{ion}} / \alpha_{\text{ES}}=1$, the ion induced currents completely cancel the ES currents and prevent pattern formation. Using the measured facet slope $m^* = \tan \ang{16}$ and assuming $m_0 = \tan \ang{25.5}$, the slope of the thermodynamically stable \{113\} facets \cite{Usui2008} (models of the facets are shown in Fig.~\ref{fig:facet}), we estimate $\alpha_{\text{ion}} / \alpha_{\text{ES}} = 0.6$. This estimate indicates that the directly ion induced currents contribute significantly to surface evolution.

\begin{figure}
	\centering
	\includegraphics{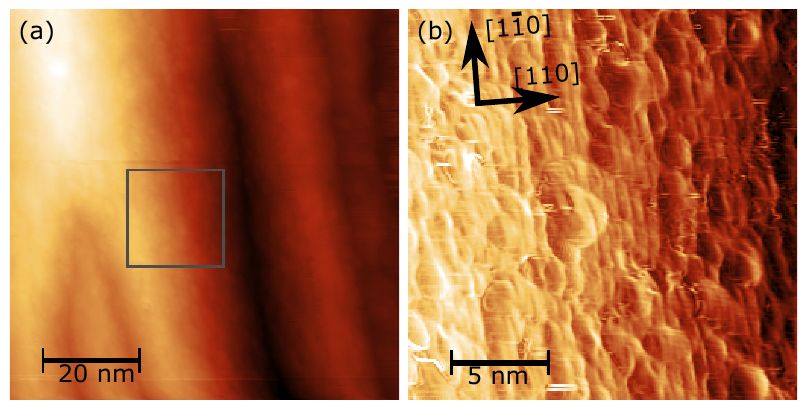}
	\caption{\label{fig:STM}
		 (a) STM image of facet on GaAs(001) after irradiation with \unit{5\E{18}}{cm^{-2}} of \unit{1}{keV} Ar at \unit{450}{\degreeC}. (b) Sobel filtered detail STM image of square area in (a).
	}
\end{figure}
Fig.~\ref{fig:STM} shows an STM image of a bifurcation (a) and a magnification (b), which is a superposition of the original topography with the Sobel filtered \cite{Hlavac2011} topography to enhance small details. The step edges in $[1\bar{1}0]$ have a separation of \unit{0.8}{nm}, corresponding to twice the distance of Ga rows on an unreconstructed Ga terminated (001) surface. On vicinal surfaces in the step flow regime the ES barrier leads to equal terrace widths \cite{Michely_Krug_2004}, as observed here. Due to the small terrace size it is reasonable to assume that the ion beam erodes the facets in step flow regime. This underlines the importance of the ES barrier for formation of large well defined facets.

Now we return to the pattern rotation observed on the back side (B) of the wafer, i.\,e. on the GaAs$(00\bar{1})$ surface. The symmetry of the GaAs lattice allows the transformation of a $(001)$ into a $(00\bar{1})$ surface by reflection at a $(001)$ plane followed by rotation by \ang{90} around $[001]$. So, the discussion above is translated to GaAs$(00\bar{1})$ just by rotating all directions by \ang{90} (compare Fig.~\ref{fig:structure}).

%\section{Conclusions}
In conclusion, we demonstrate the rotation of the ion beam induced ripple pattern on GaAs$(00 \bar{1})$ by \ang{90} compared to GaAs(001). The ripples are parallel to the preferred step edge direction $[1\bar{1}0]$ on GaAs(001) and $[110]$ on GaAs$(00\bar{1})$. 
This rotation can be explained by taking into account the symmetry of the GaAs lattice and proves that the ripples are aligned to Ga dimer rows on the Ga enriched surface and not to a specific crystallographic direction. 
We presented a model which explains this rotation.
Ion impacts produce mobile surface vacancies and adatoms. The surface currents of these species determine the development of the surface morphology. The ES barrier leads to a destabilizing surface current. Atoms displaced downhill by ion collisions form a stabilizing current counteracting the ES current and reducing the facet slope from the high symmetry surface predicted by the vanishing ES current. 
Nucleation of islands on the terraces damps short wave length fluctuation, leading to a wavelength selection. 

\begin{acknowledgement}
	The experiments were performed at the Ion Beam Center (IBC) at HZDR.
\end{acknowledgement}

\bibliography{front_back}

\end{document}